\documentclass[%
 reprint,hidelinks,nofootinbib,
 amsmath,amssymb,
 aps,
]{revtex4-2}

\usepackage{graphicx}
\usepackage{dcolumn}
\usepackage{bm}
\usepackage{mathtools}
\usepackage{csquotes}
\usepackage{hyperref}
\usepackage{float}
\usepackage{bigints}
\usepackage{color}

\begin{document}

\preprint{APS/123-QED}

\newcommand{\physrep}{{Phys. Rep.}}

\title{Magnetic dynamo caused by axions in neutron stars}

\author{Filippo Anzuini$^{1,2, 3}$}
\email{filippo.anzuini@gmail.com}
\author{Jos\'e A. Pons$^{4}$}
\author{Antonio G\'omez-Bañ\'on$^{4}$}
\author{Paul D. Lasky$^{1,2}$}
\author{Federico Bianchini$^{5,6,7}$}
\author{Andrew Melatos$^{3,8}$}

\affiliation{$^{1}$School of Physics and Astronomy, Monash University, Clayton, Victoria 3800, Australia
}
\affiliation{$^{2}$OzGrav: The ARC Centre of Excellence for Gravitational Wave Discovery, Clayton, Victoria 3800, Australia
}
\affiliation{$^{3}$School of Physics, The University of Melbourne, Parkville, Victoria 3010, Australia
}

\affiliation{$^{4}$ Departament de Física Aplicada, Universitat d'Alacant, 03690 Alicante, Spain}

\affiliation{$^{5}$ Kavli Institute for Particle Astrophysics and Cosmology, Stanford University, 452 Lomita Mall, Stanford, 
CA, 94305, USA}

\affiliation{$^{6}$
SLAC National Accelerator Laboratory, 2575 Sand Hill Road, Menlo Park, CA, 94025, USA}

\affiliation{$^{7}$ Department of Physics, Stanford University, 382 Via Pueblo Mall, Stanford, CA, 94305, USA}

\affiliation{$^{8}$ Australian Research Council Centre of Excellence for Gravitational Wave Discovery (OzGrav), The University 
of Melbourne, Parkville, Victoria 3010, Australia}

\date{\today}
\newcommand{\jcap}{{JCAP}}
\newcommand{\aap}{{A\&A}}
\newcommand{\mnras}{{MNRAS}}
\newcommand{\apjl}{{ApJ,}}
\newcommand{\apjs}{{ApJS,}}
\begin{abstract}
The coupling between axions and photons modifies Maxwell's equations, introducing a dynamo term in the magnetic induction equation. In neutron stars, for critical values of the axion decay constant and axion mass, the magnetic dynamo mechanism increases the total magnetic energy of the star. We show that this generates substantial internal heating due to enhanced dissipation of crustal electric currents. These mechanisms would lead magnetized neutron stars to increase their magnetic energy and thermal luminosity by several orders of magnitude, in contrast to observations of thermally-emitting neutron stars. To prevent the activation of the dynamo, bounds on the allowed axion parameter space can be derived.
\end{abstract}

\maketitle


\textit{Introduction.}--Axions were originally introduced to solve the strong CP problem \citep{Peccei_1977a, Peccei_1977b, Wilczek_1978, Weinberg_1978, Cortona_2016, Di_Luzio_2020}. Current constraints on axions are based on a wide range of techniques, from laboratory experiments to astrophysics \citep{SQUID_2010, Budker_2014, Arvanitaki_2014, Kahn_2016, CAST_2017, Irastorza_2018, Di_Luzio_2020, Di_Luzio_2022b, Dolan_2022}. Bounds on the axion parameter space are obtained using helioscopes or haloscopes \citep{Sikivie_1983, Sikivie_1984, Hagmann_1990, SQUID_2010, Du_2018, Knirck_2019, Kim_2019, Braine_2020}, calculating axion emission in neutron stars (NSs) \citep{Sedrakian_2016, Hamaguchi_2018,Sedrakian_2019, Buschmann_2021}, and with other astrophysical and cosmological constraints \citep{Raffelt_2008, Leinson_2014, Marsh_2016, Garbrecht_2018, Day_2019, Dolan_2021, Di_Luzio_2022a, Balazs_2022}. 

It has been recently shown that finite density corrections change the axion field in NSs, which can become large compared to the vanishing expectation value in vacuo \citep{Hook_2018, Huang_2019, Balkin_2020, Di_Luzio_2021}, and may mediate long-range forces in binary NS systems \citep{Hook_2018, Huang_2019, Zhang_2021}. Under these conditions, additional observable effects can be used to constrain the axion parameter space.

In this Letter, we take advantage of these and other astrophysical aspects of NSs. The ultra-strong magnetic fields of NSs allow us to test axion electrodynamics beyond the strengths attainable in terrestrial laboratories. The axion-photon coupling introduces additional terms in Maxwell's equations \cite{Wilczek_1987, Visinelli_2013, Kim_2019}. In this work we highlight
that the such coupling introduces a dynamo term in the magnetic induction equation, which can be used to place constraints on axion parameters, and to test models of axions in dense matter. By studying the specific case of crust-confined magnetic fields, we find that when the mass and decay constant of the axion reach critical thresholds, the dynamo supplies the star with additional magnetic energy. As the magnetic field gets stronger, the stellar interior is heated up by enhanced dissipation of electric currents in the crust. This process increases the observable thermal luminosity, in tension with observations of thermally-emitting, isolated NSs for a wide range of parameters. 

To reconcile the presence of axions in the interior of NSs with observations, we appeal to one of the following scenarios:
(i) the axion mass, decay constant, and axion-photon coupling constant (that determine the axion field profile in dense matter and regulate the interaction with electromagnetic fields) are below the critical values that activate the axion dynamo in the magnetized crust; (ii) a nonlinear saturation mechanism may halt the supply of magnetic energy via the dynamo beyond a certain threshold; or (iii) NSs do not source axions \citep{Hook_2018, Huang_2019, Balkin_2020, Zhang_2021} (or they do not couple with photons).

\textit{Maxwell-axion equations.}-- Two sets of Maxwell's equations with axions are used in the literature: the set preserving the duality symmetry \citep{Visinelli_2013, Ko_2016, Tercas_2018}, and the \enquote{standard} set, often used in direct axion searches \citep{Sikivie_1983, Wilczek_1987, Kim_2019}. In this Letter, we employ the Maxwell-axion equations derived in \citep{Visinelli_2013}. The practical advantage is that the dynamo term appears explicitly in the magnetic induction equation. We simulate the magneto-thermal evolution of NSs with an active axion-dynamo, for highly and weakly magnetized stars (for the latter, see the Supplemental material, Section \ref{sec:Weakly_mag_NS}) and for different initial magnetic configurations (cf. Sections \ref{sec:Maps} and \ref{sec:Core_fields} in the Supplemental material). The dynamo term is still present in the standard set of Maxwell's equations \citep{Sikivie_1983, Wilczek_1987, Long_2015, Rogachevskii_2017}, which we study in the Supplemental material (Section \ref{sec:Maxwell_equations_setII}) following \citep{Kim_2019}. 

We use an interior metric corresponding to a spherically symmetric, non-rotating body, with the notation
$ds^2 = -e^{2 \nu(r)} c^2 dt^2 + e^{2 \lambda(r)} dr^2 + r^2d\Omega^2$.
In the absence of magnetic monopoles and magnetic currents, Maxwell's equations read \citep{Visinelli_2013, Ko_2016, Tercas_2018}

\begin{subequations}
\begin{align}
   &  \nabla \cdot (\mathbf{E} - g_{a\gamma}a\mathbf{B}) = 4\pi\rho_c \\
   \nonumber \\
   & \nabla \times [e^\nu(\mathbf{B} + g_{a\gamma}a\mathbf{E})] = \frac{\partial (\mathbf{E} - g_{a\gamma}a\mathbf{B})}{c \ \partial t} +  \frac{4\pi e^\nu}{c}\mathbf{J}  \ \ \\
    \nonumber\\
    &  \nabla \cdot (\mathbf{B} + g_{a\gamma}a\mathbf{E}) = 0  \\
   \nonumber \\
   & \nabla \times [e^\nu(\mathbf{E} - g_{a\gamma}a\mathbf{B})] = -\frac{1}{c}\frac{\partial (\mathbf{B} + g_{a\gamma}a\mathbf{E})}{\partial{t}} \ .
   \end{align}
   \label{eq:VIS_eqs}
   \end{subequations}
In our notation, the $\nabla$ operator includes the metric factors. Here, $a$ is the axion field, $\rho_c$ denotes the charge density, and $\mathbf{J}$ is the electric current density. 
The axion-photon coupling is
$g_{a\gamma} = g_{\gamma}\alpha_{\textrm{EM}}/\pi f_a$, where $f_a$ is the axion decay constant, $g_{\gamma} = 0.37$ \citep{Cortona_2016}, and $\alpha_{\textrm{EM}}$ denotes the fine-structure constant.

Since we are interested in the long-term evolution on timescales of thousands to millions of years, and in the highly conductive NS interior $E \approx (v/c) B$ (where $1 \ \textrm{km}/\textrm{Myr} \lesssim v \lesssim 10^3 \ \textrm{km}/\textrm{Myr}$ is the electron drift velocity in the crust), one can adopt some simplifying assumptions.
In Eq. (\ref{eq:VIS_eqs}b), we neglect the term proportional to $\partial \mathbf{E}/\partial t$ (standard magneto-hydrodynamics approximation) and $g_{a\gamma}\partial(a\mathbf{B})/\partial t$, which is typically much smaller than $\mathbf{J}$. We also use $\nabla\times(\mathbf{B} + g_{a\gamma}a\mathbf{E}) \approx \nabla\times\mathbf{B}$. In the induction equation, we neglect $g_{a\gamma}\partial(a\mathbf{E})/\partial t$. We check numerically that the terms above contribute with small or negligible corrections to the magnetic field evolution. We use the simplified version of Maxwell's equations below

\begin{subequations}
\begin{align}
   &\nabla \cdot (\mathbf{E} - g_{a\gamma}a\mathbf{B}) = 4\pi\rho_c \\
   \nonumber \\
   &\nabla \times (e^\nu\mathbf{B}) = \frac{4\pi e^\nu}{c}\mathbf{J}  \ \\
    \nonumber\\
   &\nabla \cdot \mathbf{B} = 0 \\
   \nonumber \\
   &\nabla \times [e^\nu(\mathbf{E} - g_{a\gamma}a\mathbf{B})] = -\frac{1}{c}\frac{\partial \mathbf{B}}{\partial{t}} \ .
   \label{eq:Maxwell_eq_induction}
   \end{align}
   \label{eq:Maxwell_eqs_approx}
   \end{subequations}
The system is closed with the electric field derived from a generalized Ohm's law \citep{Pons_2019}
\begin{equation}
    \mathbf{E} = \frac{\mathbf{J}}{\sigma_e} + \frac{\mathbf{J} \times  \mathbf{B}}{e n_e c} \ ,
    \label{eq:e_field}
\end{equation}
where $\sigma_e$ denotes the electric conductivity, while $e$ and $n_e$ denote the electric charge and the electron number density respectively. Eq. \eqref{eq:e_field} is also employed in previous works on axions \citep{Long_2015, Rogachevskii_2017, Tercas_2018}. We expect $\sigma_e$ to remain in the same range as in the case without axions, i.e. $\sigma_e \in [10^{22}, 10^{29}]$ s$^{-1}$ (see Figure 1 in \citep{Pons_2007} for the variation of $\sigma_e$ in the crust). In such range, $\mathbf{E} \lesssim 10^{-15} \mathbf{B}$, which justifies the omission of terms such as $g_{a\gamma}\nabla \times (a\mathbf{E})$. Inserting the generalized Ohm's law in the induction equation, we get 

\begin{eqnarray}
\frac{\partial \mathbf{B}}{\partial t} = -\boldsymbol{\nabla}\times \Big[ && \eta \boldsymbol{\nabla}\times (e^{\nu}\mathbf{B}) + f_H[\boldsymbol{\nabla} \times (e^{\nu}\mathbf{B})] \times \mathbf{B}  \nonumber\\
&& - g_{a\gamma}cae^{\nu}\mathbf{B}\Big] \ ,  
\label{eq:mag_dynamo}
\end{eqnarray}
where $\eta = c^2/(4\pi\sigma_e)$ is the magnetic diffusivity and $f_H = c/(4\pi e n_e)$.
The first term in Eq. \eqref{eq:mag_dynamo} accounts for ohmic dissipation, and the second term for the Hall drift. The critical effect of axions is to include the third, {\it dynamo-like} term. It may be interpreted as a dynamo or anti-dynamo, depending on the overall sign, and can increase or decrease, locally, of the magnetic field strength (see the Supplemental material for the case of the standard axion-Maxwell equations \citep{Sikivie_1983, Kim_2019}).
The activation of the dynamo is regulated by the interplay of the ohmic dissipative term, the Hall term, the magnitude of $g_{a\gamma}$ and of the axion gradient $\nabla a$.
Dynamos (without axions) also appear when simulating the magnetic field amplification in proto-NSs \citep{Thompson_1993, Bonanno_2005, Naso_2008, Reboul_2021}, such as in the case of the magneto-rotational instability with the $\alpha-\Omega$ dynamo, or due to magnetized convection.

\textit{Axions in dense matter.}--We now review the case of NSs sourcing axions. In \citep{Hook_2018, Huang_2019, Balkin_2020, Di_Luzio_2021} it is shown that, under certain conditions, the in-medio axion potential has minima at $a/f_a \approx \pm \pi$, contrarily to the in-vacuo potential, which is minimized at $a/f_a \approx 0$.  The axion field profile in dense matter is the solution of the Klein-Gordon (KG) equation
\begin{eqnarray}
    \Box a - \frac{\partial V}{\partial a} =  \frac{g_{a\gamma}}{4\pi}\mathbf{E} \cdot \mathbf{B} \ ,
\end{eqnarray}
where $V$ is the in-medio axion potential. In vacuo, the axion potential reads \citep{Cortona_2016, Di_Luzio_2021}
\begin{eqnarray}
V_0(a) = -m_\pi^2 f_\pi^2 \sqrt{1 - \frac{4m_um_d}{(m_u+m_d)^2}\sin^2\left(\frac{a}{2f_a}\right)} \ , 
\label{eqn:axion_potential_vac}
\end{eqnarray}
where $m_u$ and $m_d$ are respectively the up and down quark masses, and $m_{\pi} = 135$ MeV and $f_{\pi} = 93$ MeV denote the mass of the neutral pion and the pion decay constant respectively. Finite density corrections modify the axion potential, which in the linear approximation reads \citep{Hook_2018, Balkin_2020, Di_Luzio_2021}

\begin{eqnarray}
V(a) \approx \left(1 - \frac{\sigma_N n_N}{m^2_\pi f^2_\pi} \right)  V_0(a) \ , \ \ \ \
\label{eqn:axion_potential}
\end{eqnarray}
where $\sigma_\textrm{N} = 59$ MeV and $n_\textrm{N}$ is the nucleon number density. Different forms of the in-medio axion potential have been derived \citep{Hook_2018, Huang_2019, Balkin_2020, Zhang_2021,Di_Luzio_2021}. We rely on the simple potential in Eq. \eqref{eqn:axion_potential} exclusively to determine where the shift of the axion field takes place (see below). This occurs above the nuclear saturation density, which may be caused by exotic phases in dense matter \citep{Balkin_2020}.

The key point in \citep{Hook_2018, Huang_2019, Balkin_2020, Zhang_2021} is that the axion field is shifted to $a/f_a \approx \pm \pi$ at densities above a critical value, set by the condition that the prefactor
$(1 - \sigma_\textrm{N}n_\textrm{N}/m^2_\pi f^2_\pi)$
vanishes in the NS interior \citep{Hook_2018, Huang_2019, Balkin_2020, Di_Luzio_2021}. We denote the corresponding critical radius with $r = r_{\textrm{crit}}$.
For $r > r_{\textrm{crit}}$, the axion field decreases exponentially on a length-scale of the order of the inverse axion mass in vacuo $m_a^{-1}$ \citep{Huang_2019, Balkin_2020}, with
$m_a = m_{\pi} f_\pi\sqrt{z}/[f_a(1+z)]$ (where $z = m_u/m_d \approx 0.48$ \citep{Cortona_2016, Di_Luzio_2021}).

Typically one has $10^8 \ \textrm{GeV} \lesssim f_a \lesssim 10^{18}$ GeV and $10^{-11} \ \textrm{eV} \lesssim m_a \lesssim 10^{-2}$ eV \citep{Irastorza_2018, DiLuzio_2020} for the QCD axion. For the parameters employed here ($f_a \approx 10^{15}$ GeV and $m_a \approx 10^{-9}$ eV) and typical NS magnetic field strengths, the electromagnetic backreaction on the axion field in the KG equation is negligible compared to the potential term, and the latter gives a stiff source. The solution to the KG equation is a field that oscillates with a frequency close to its Compton frequency, i.e. on a time-scale far shorter than the typical, secular time-scales for the NS magneto-thermal evolution. On such long time-scales, only the axion profile time-averaged over several oscillations plays a role, and it shows negligible deviations from the stationary profile ($\lesssim$ few parts in $10^{4}$). We approximate the latter as in \citep{Balkin_2020}, i.e.

\begin{eqnarray}
   a(r) \approx
   \begin{cases}
    \pm \pi f_a \qquad \qquad \qquad \qquad \ \ \ r \leq r_{\textrm{crit}}\\
    \\
    \pm \pi f_a \frac{r_{\textrm{crit}}}{r} e^{-m_a(r-r_{\textrm{crit}})} \ \ \ \ \ \ r > r_{\textrm{crit}} \ .
   \end{cases}
   \label{eq:axion_stationary}
\end{eqnarray}
Below we employ $a/f_a = \pi$ at high densities. We stress that the numerical results presented in this work rely on Eq. \eqref{eq:axion_stationary}, and on the values of $f_a, m_a$ and $r_{\textrm{crit}}$.  

The potential shift $\Delta V$ due to the sourcing of the axion field is $\Delta V \approx (m_\pi f_\pi)^2$ (for $a \approx \pi f_a$) \citep{Hook_2018, Balkin_2020}, and allows us to determine an upper limit to the magnetic energy that the dynamo can convert, which amounts approximately to $\lesssim 10 \%$ of the NS binding energy \citep{Lattimer_2001}, and exceeds by orders of magnitude the magnetic energy expected in magnetars. However, for energies comparable to those of magnetars ($\approx 10^{48}$-$10^{49}$ erg), the backreaction on the axion field and non-linear saturation mechanisms may become important and may halt the growth of the magnetic energy.

The dynamo is active as long as $\nabla a$ is maintained in the magnetized region of the star, and operates on a time-scale $\tau_{\rm{dyn}} \approx |g_{a\gamma}c\nabla a|^{-1}$ (cf. Eq. \eqref{eq:mag_dynamo}). Given the axion exponential profile in Eq. \eqref{eq:axion_stationary}, $\tau_{\rm{dyn}}$ varies by several orders of magnitude, depending on $f_a$. As shown below, the dynamo can make an observable difference for isolated NSs when $\tau_{\rm{dyn}}$ is shorter than the typical cooling time-scale ($\sim 1$ Myr), which in the crust occurs for $f_a \gtrsim 3.6 \times 10^{15}$ GeV.

\begin{figure*}
\includegraphics[width=17.5cm, height = 6cm]{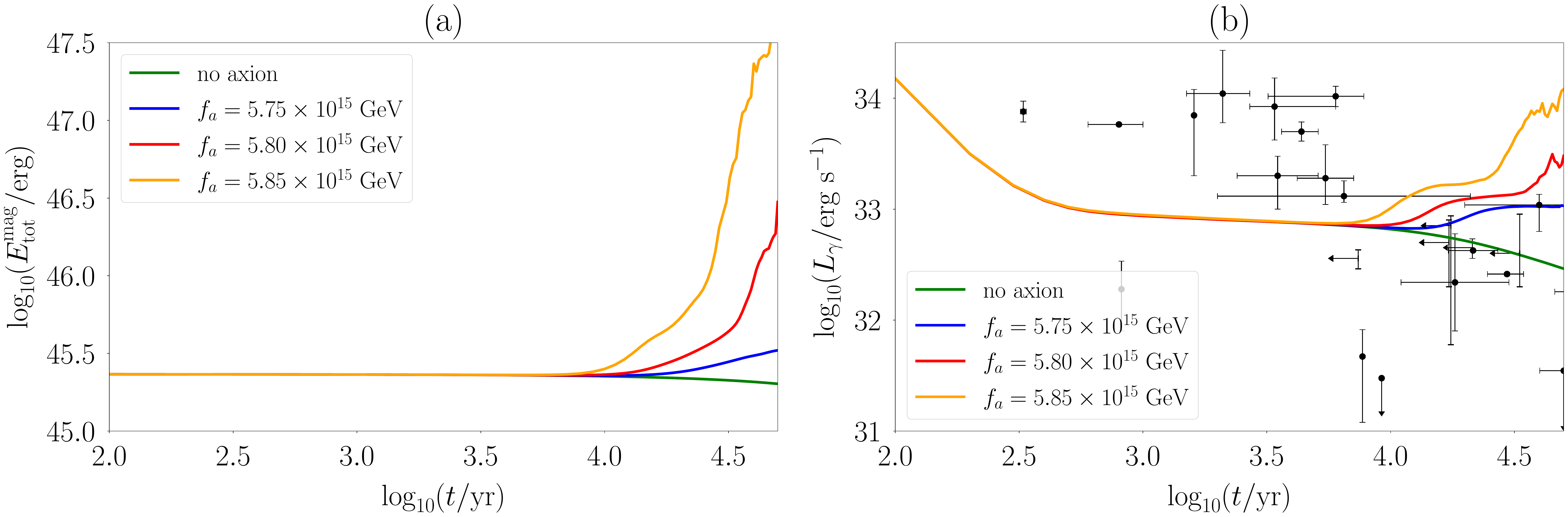}
\caption{\label{fig:Cool_curves_energy_main} Magnetic energy (panel (a)) and redshifted thermal luminosity $L_{\gamma}$ (panel (b)) versus stellar age for different values of $f_a$ and for the A2 initial magnetic configuration (cf. Table \ref{tab:table_1}). The black points correspond to thermally-emitting, isolated NSs with inferred fields below $10^{14}$ G (taken from \citep{Potekhin_2020}).}
\end{figure*}

\begin{table}[b]
\caption{\label{tab:table_1}%
Initial magnetic field configurations considered in this work. The polar surface strength of the dipolar-poloidal field is $B_{\textrm{dip}}$. $E^{\textrm{mag}}_{\textrm{tor}}$ and $E^{\textrm{mag}}_{\textrm{tot}}$ are the toroidal and total magnetic energy respectively. $l_{\textrm{pol}}$ denotes the number of poloidal multipoles.
All the magnetic configurations are crust-confined, except for the C1 configuration. The latter includes also a toroidal field in the core region confined to an equatorial torus.}
\begin{ruledtabular}
\begin{tabular}{lccr}%
\textrm{Config.}&
$B_{\textrm{dip}}$&
\multicolumn{1}{c}{$E^{\textrm{mag}}_{\textrm{tor}}/E^{\textrm{mag}}_{\textrm{tot}}$}&
$l_{\textrm{pol}}$\\
\colrule

A0 & $1.0 \times 10^{10}$ G & $ 93\%$ & $1$ \\
A1m$_{2,2}$ & $1.0 \times 10^{13}$ G & $77\%$ & $2$  \\
A1 & $5.0 \times 10^{13}$ G & $ 35\%$ & $1$ \\
A2 & $3.0 \times 10^{13}$ G & $ 0$ & $1$ \\
C1 & $2.0 \times 10^{13}$ G & $43 \%$ & $1$\\

\end{tabular}
\end{ruledtabular}
\end{table}

\textit{The dynamo mechanism.}--We employ the two-dimensional, axisymmetric magneto-thermal evolution numerical code for NSs developed by the Alicante group  \citep{Aguilera_2008, Pons_2009, Vigano_2012, Vigano_2013, Pons_2019, Dehman_2020, Vigano_2021}, recently modified to include hyperon species (GM1A equation of state \citep{Gusakov_2014}) in \cite{Anzuini_2021, Anzuini_2022}. We consider a NS with mass $M = 1.8 \ M_{\odot}$ with superfluid nucleons and hyperons, with electrons, muons and an iron-only outer envelope (cf. \citep{Anzuini_2021, Anzuini_2022} for details). In principle, modifications to the equation of state and several microphysical details are affected by a large axion field. However, the magnetic evolution is insensitive to such changes. Indeed, NS models obtained with different equations of state without axions (and hence with different structural and compositional properties, cf. for example \citep{Dehman_2020, Vigano_2021} with \citep{Anzuini_2021, Anzuini_2022}) have a similar magnetic evolution, and we expect the same even in the presence of axions. Hence, we include the effect of axions only via Maxwell's equations. Below, we focus on the A2 initial, crust-confined magnetic configuration, with a dipolar-poloidal, quadrupolar-toroidal field (see Table \ref{tab:table_1}). Other configurations (both crust-confined and core-threading) are considered in the Supplemental Material.

The total magnetic energy $E^{\textrm{mag}}_{\textrm{tot}}$ and the redshifted thermal luminosity $L_{\gamma}$ of the NS model are displayed in Figure \ref{fig:Cool_curves_energy_main}. Panel (a) reports the evolution of $E^{\textrm{mag}}_{\textrm{tot}}$ for stars sourcing axions (blue, red and orange curves) and without axions (green curve). To improve the stability of the simulations, we choose $f_a$ such that $E^{\textrm{mag}}_{\textrm{tot}}$ grows on secular time-scales. The dynamo supplies magnetic energy in the crust at a faster pace than the dissipative processes can reduce it. The higher is $f_a$, the shorter is $\tau_{\textrm{dyn}}$. For the orange curve, $E^{\textrm{mag}}_{\textrm{tot}}$ is two orders of magnitude higher than the model without axions after $10^{4.5}$ yr.

Figure \ref{fig:Cool_curves_energy_main}(b) shows the corresponding cooling curves. NSs without axions match the data of thermally-emitting, isolated NSs with inferred fields $\lesssim 10^{14}$ G at $\gtrsim 10^4$ yr \citep{Potekhin_2020} (stronger initial fields match younger sources, cf. the Supplemental material). In NSs sourcing axions, the dissipation of intense electric currents that sustain the growing magnetic field enhances the Joule heating rate, increasing $L_{\gamma}$. In general, models without axions overlap with the data for $\lesssim 1$ Myr (see \citep{Anzuini_2021, Anzuini_2022}); models with axions overshoot the observed $L_{\gamma}$ of several sources with $L_{\gamma} \lesssim 10^{33}$ erg/s. We expect however $L_{\gamma}$ to saturate to a certain level, set by the balance between Joule heating and cooling processes ($>10^{33}$ erg/s for the parameters employed here).

\begin{figure*}
\includegraphics[width=18cm, height = 10cm]{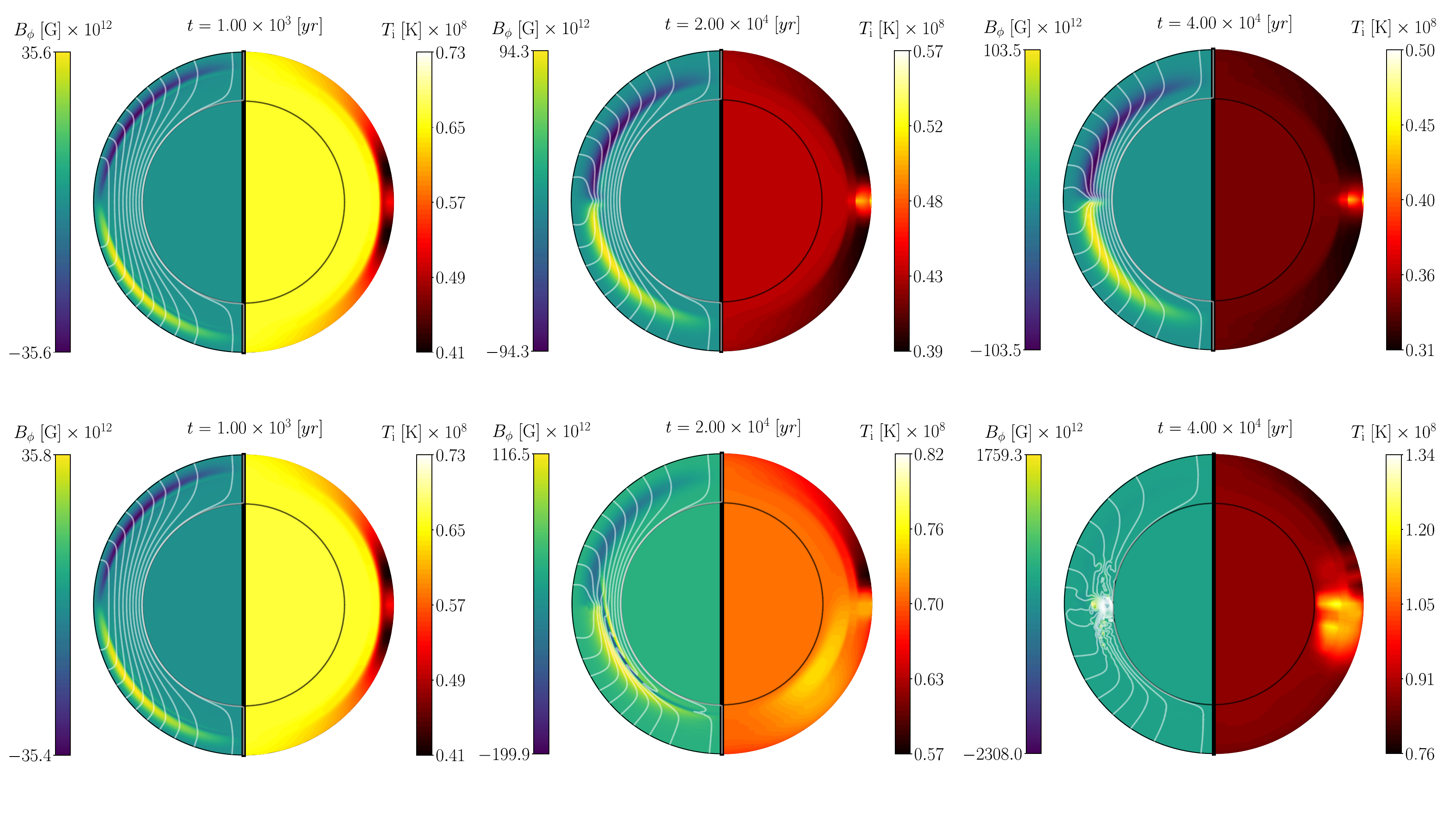}
\caption{Magnetic and temperature maps for the A2 configuration. The top panels correspond to a star that does not source axions; the bottom panels to a star sourcing an axion field with $f_a = 5.8 \times 10^{15}$ GeV. The maps extend from the NS center to the bottom of the outer envelope, i.e. the latter is not included in the plots. The crust is enlarged by a factor $8$ for visualization purposes. The scales in the colorbars change in time to preserve a high level of detail.}
\label{fig:Pol_only_main}
\end{figure*}

In Figure \ref{fig:Pol_only_main} we display the magneto-thermal maps of a NS without axions (top panels) and with axions for $f_a = 5.8 \times 10^{15}$ GeV (bottom panels). The left hemispheres show the contours of the toroidal field $\mathbf{B}_\textrm{tor} = B_{\phi}\hat{\phi}$. Overplotted are the poloidal field lines (projected onto the meridional plane). The right hemispheres display the redshifted internal temperature $T_{\textrm{i}}$. The dynamo mechanism produces rich dynamics in the bottom panels. At $t = 4\times 10^4$ yr, $B_{\phi}$ is approximately one order of magnitude stronger for a star with axions (bottom panel) than for a star without axions (top panel). The temperature map shows that enhanced Joule heating due to the dynamo raises $T_{\textrm{i}}$ at the equator with respect to the top panel.

We stress that, due to unavoidable theoretical and numerical limitations \citep{Braithwaite_2006, Lander_2009, Gourgouliatos_2013, Gourgouliatos_2016, Wood_2015, Igoshev_2021}, there is limited capability to reproduce realistic NS magnetic field configurations, where small-scale fields (which enhance ohmic dissipation) may be endemic. However, a slight increase of $f_a$ promotes the efficiency of the dynamo, so that the $f_a$ values studied here should remain valid even with more complex configurations. We also note that such values of $f_a$ allow us to test $g_{a\gamma} \approx 10^{-19}$ GeV$^{-1}$, which is orders of magnitude smaller than the sensitivity of ADMX ($g_{a\gamma} \approx 10^{-16}$ GeV$^{-1}$ \citep{Braine_2020}).

\textit{Conclusion} -
NSs are ideal environments to test axion electrodynamics, and can be used to derive constraints on the axion parameter space, complementing ground-based axion searches \citep{SQUID_2010, Budker_2014, CAST_2017, Kim_2019, Braine_2020}, and to assess proposed axion models \citep{Hook_2018, Huang_2019, Balkin_2020, Zhang_2021, Di_Luzio_2021}.

We find that the coupling of axions with photons leads to the appearance of a magnetic dynamo term in the magnetic induction equation. In NSs sourcing axions with $a\approx \pm \pi f_a$ \citep{Hook_2018, Huang_2019, Balkin_2020}, the magnetic energy of the star increases by orders of magnitude, for certain threshold values of the axion decay constant and axion mass. The thermal luminosity $L_{\gamma}$ increases by roughly one order of magnitude with respect to NSs without axions due to enhanced dissipation of the strong electric currents that sustain the growing magnetic field. Values of $L_{\gamma}$ in the range $\gtrsim 10^{33}$ erg/s maintained by the dynamo are in tension with available data, in particular with mature NSs with ages $\gtrsim 10^4$ yr and $L_{\gamma} \lesssim 10^{33}$ erg/s \citep{Potekhin_2020}.

The activation of the dynamo may be prevented under certain conditions. It does not activate if $f_a$ and $m_a$ are such, that the dynamo term is sufficiently small in the magnetized crust. Moreover, one cannot exclude a priori an efficient mechanism, which may saturate the magnetic energy of the star beyond a certain threshold. Another possibility is that the axion field is not shifted to $a \approx \pm \pi f_a$ in NS interiors \citep{Hook_2018, Huang_2019, Balkin_2020, Zhang_2021, Di_Luzio_2021}. 

We conclude with some caveats. The results presented in this work are calculated using the axion profile in Eq. \eqref{eq:axion_stationary}, which is an approximate solution to the KG equation. Moreover, in realistic NS magnetic field configurations, ohmic dissipation may be enhanced due to the presence of small-scale fields. However, a slight change in $f_a$ increases the efficiency of the dynamo, so that the range of $f_a$ studied in this Letter is expected to remain similar even in the presence of more complex configurations.

\begin{acknowledgments}
FA thanks Koichi Hamaguchi, Reuven Balkin, Hugo Ter\c{c}as, David J. E. Marsh and Angelo Maggi for valuable discussions. FA, PDL and AM are supported by the Australian Research Council (ARC) Centre of Excellence for Gravitational Wave Discovery (OzGrav), through project number CE170100004. PDL is supported through ARC Discovery Project DP220101610. JAP acknowledges support by the Generalitat Valenciana (PROMETEO/2019/071) and AEI grant PID2021-127495NB-I00.
\end{acknowledgments}

\bibliography{apssamp}
\newpage
\onecolumngrid
\begin{center}
\textbf{\large Supplemental Material}
\end{center}
\setcounter{equation}{0}
\setcounter{figure}{0}
\makeatletter
\renewcommand{\theequation}{S\arabic{equation}}
\renewcommand{\thefigure}{S\arabic{figure}}

The following material studies the evolution of the total magnetic energy of a star with a weak initial magnetic field when the axion dynamo is active (Section \ref{sec:Weakly_mag_NS}), and provides a detailed study of the magneto-thermal evolution of NSs with different magnetic topologies from the one studied in the main Letter (Sections \ref{sec:Maps} and \ref{sec:Core_fields}). In Section \ref{sec:Maxwell_equations_setII} we derive the dynamo for the \enquote{standard} Maxwell-axion equations \citep{Sikivie_1983, Wilczek_1987} commonly used in direct axion searches, following the effective approximation presented in \cite{Kim_2019}.

\section{Dynamo effect in weakly magnetized stars}
\label{sec:Weakly_mag_NS}

\begin{figure}[!h]
\includegraphics[width=8cm, height = 5.5cm]{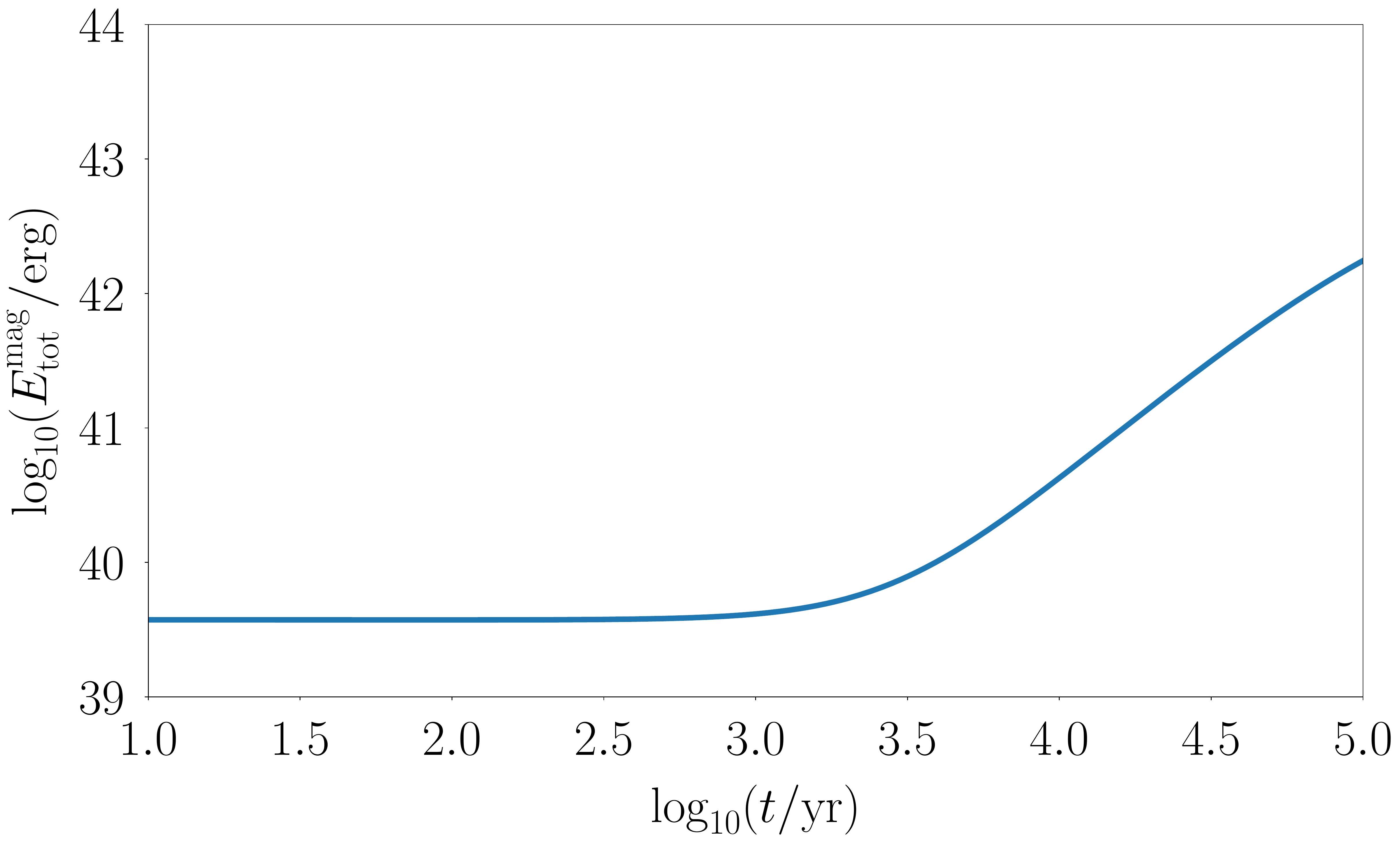}
\caption{\label{fig:Weakly_mag_star} Evolution of $E^{\textrm{mag}}_{\textrm{tot}}$ for a weakly magnetized star (A0 configuration in Table \ref{tab:table_1}) sourcing axions, with $f_a = 5.98 \times 10^{15}$ GeV.}
\end{figure}

In this section we show that if the axion dynamo is active, even stars with relatively weak fields can increase their total magnetic energy by several orders of magnitude. Figure \ref{fig:Weakly_mag_star} studies the evolution of the total magnetic energy $E^{\textrm{mag}}_{\textrm{tot}}$ for the A0 configuration, with a dipolar-poloidal and quadrupolar-toroidal magnetic field (with strength of the poloidal-dipolar magnetic field component at the polar surface given by $B_{\textrm{dip}} = 10^{10}$ G and $E^{\textrm{mag}}_{\textrm{tor}}/ E^{\textrm{mag}}_{\textrm{tot}} = 93 \%$ at $t = 0$, where $E^{\textrm{mag}}_{\textrm{tor}}$ denotes the magnetic energy in the toroidal component, cf. Table \ref{tab:table_1}), for $f_a = 5.98 \times 10^{15}$ GeV. The comparison of $E^{\textrm{mag}}_{\textrm{tot}}$ at $t = 0$ and at $t = 10^5$ yr shows that $E^{\textrm{mag}}_{\textrm{tot}}$ increases by more than two orders of magnitude.

We conclude that the axion dynamo can increase the initial magnetic energy budget even in weakly magnetized stars, which would evolve eventually into highly magnetized NSs.\\

\section{\label{sec:Maps}Cooling curves and magneto-thermal maps}

\begin{figure*}[!h]
\includegraphics[width=17.5cm, height = 6cm]{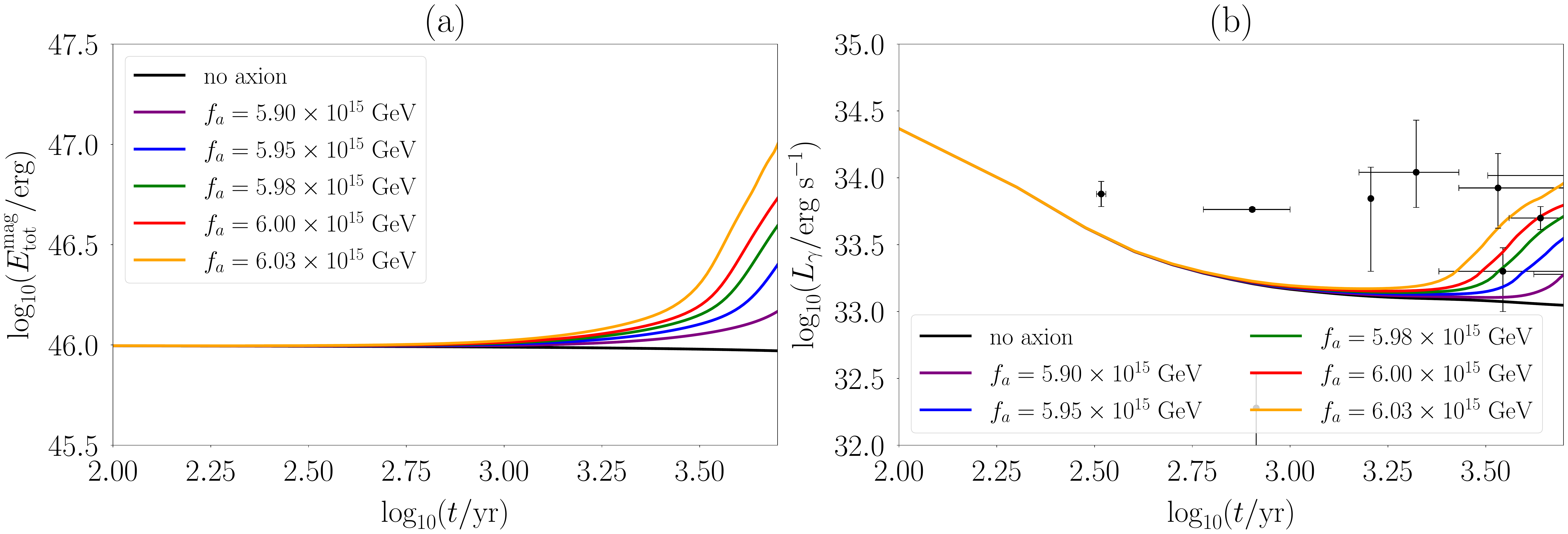}
\caption{\label{fig:Cool_curves_energy} Magnetic energy evolution and cooling curves for the A1 initial magnetic configuration, (see Table \ref{tab:table_1}). \textit{(a)} Total magnetic energy of the star $E^{\textrm{mag}}_{\textrm{tot}}$ versus stellar age. \textit{(b)} Redshifted thermal luminosity $L_{\gamma}$ versus stellar age. When the dynamo is active, $L_{\gamma}$ grows on a time-scale set by $f_a$. The higher $L_{\gamma}$ becomes, the harder it is to match the thermal luminosity data of isolated NSs with ages above $10^4$ yr (cf. Figure \ref{fig:Cool_curves_energy_main} in the main Letter).}
\end{figure*}

We now extend the results presented in the main Letter and perform a set of magneto-thermal simulations for different initial magnetic field configurations.
First, we consider a mixed dipolar-poloidal, quadrupolar-toroidal crust-confined magnetic field, with $B_{\textrm{dip}} = 5 \times 10^{13}$ G and with $E^{\textrm{mag}}_{\textrm{tor}}/ E^{\textrm{mag}}_{\textrm{tot}} = 35 \%$ at $t = 0$ (configuration A1 in Table \ref{tab:table_1}). Figure \ref{fig:Cool_curves_energy}(a) displays $E^{\textrm{mag}}_{\textrm{tot}}$ versus the stellar age. For $f_a \gtrsim 5.90 \times 10^{15}$ GeV, the total magnetic energy of the star increases over time. At $t = 5\times 10^3$ yr and for $f_a = 6.03 \times 10^{15}$ GeV, $E^{\textrm{mag}}_{\textrm{tot}}$ is more than one order of magnitude higher than for a star that does not source axions. 

\begin{figure*}[!h]
\includegraphics[width=18cm, height = 10cm]{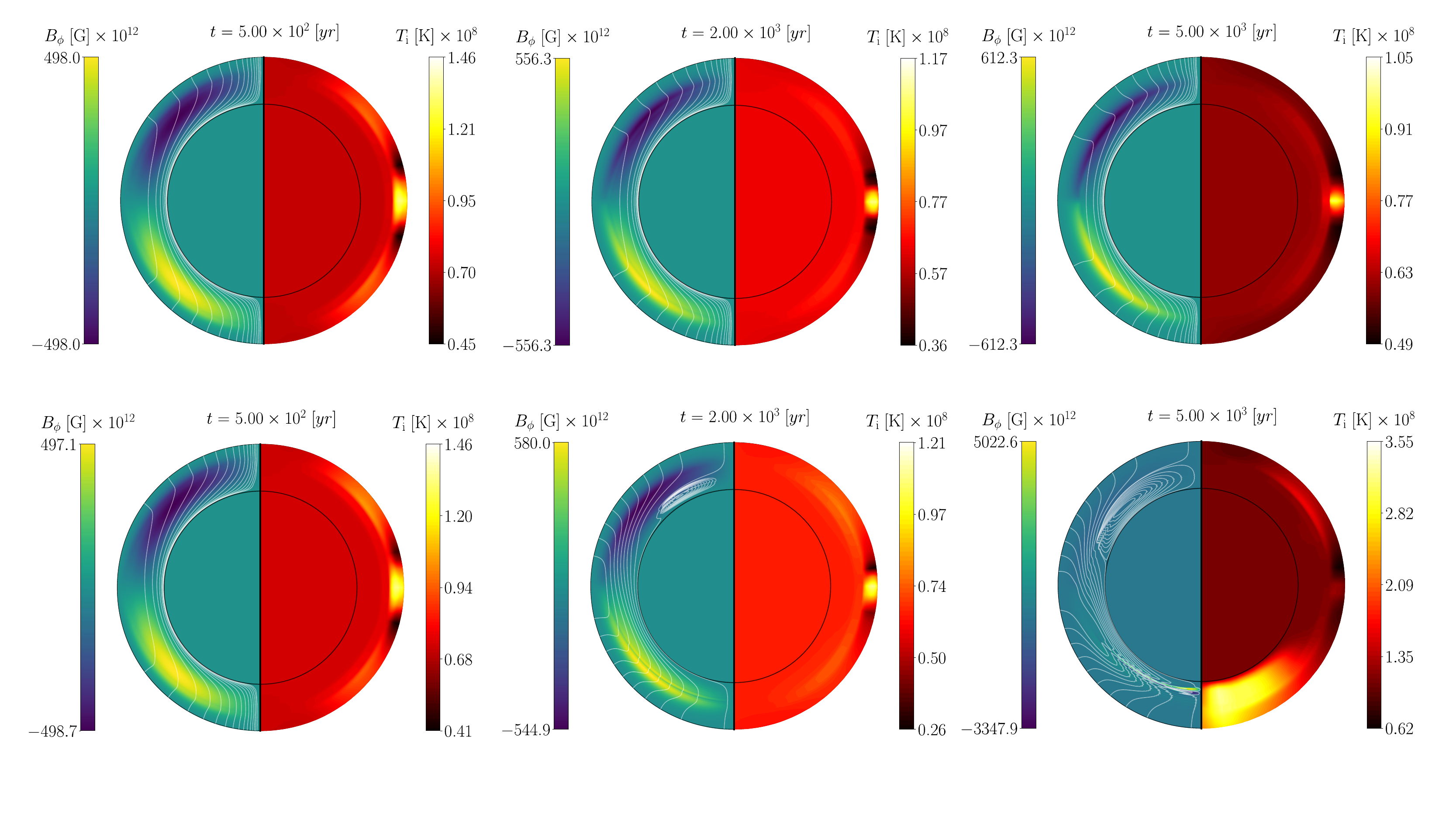}
\caption{Magneto-thermal evolution for the A1 initial configuration (see Table \ref{tab:table_1} for details). The contours of the toroidal field $B_{\phi}$ are reported in the left hemispheres. The overplotted white lines represent the projections onto the meridional plane of the poloidal field lines. The right hemispheres display $T_{\textrm{i}}$, the redshifted internal temperature. In the top panels, the snapshots correspond to a NS without axions; the bottom panels to a star sourcing an axion field ($f_a = 5.98 \times 10^{15}$ GeV). As in the main Letter, the crust is enlarged by a factor $8$ for visualization purposes, and the scales of the colorbars change between the different snapshots to maintain a high level of detail.}
\label{fig:pol_tor}
\end{figure*}

The cooling curves for the different models are reported in Figure \ref{fig:Cool_curves_energy}(b). The dynamo mechanism generates strong electric currents, leading to higher Joule heating rates than for stars not sourcing axions, causing the star to attain higher values of $L_{\gamma}$. The purple curve is similar to the black curve (model not sourcing axions), and is visibly higher than the black curve only for $t \gtrsim 3 \times 10^3$ yr. However, the values of $L_{\gamma}$ attained by the other curves grow faster; in particular, $L_{\gamma}$ reached by the orange curve at $t = 5\times 10^3$ yr exceeds by almost one order of magnitude the corresponding one for the black curve. The curves match some of the data of relatively young, thermally-emitting and isolated NSs (taken from \citep{Potekhin_2020}). At later times however, the growth of $L_{\gamma}$ is in tension with the observed thermal luminosity of sources with $L_{\gamma} \lesssim 10^{33}$ erg/s (cf. Figure 1 in the main Letter).

\begin{figure*}[!h]
\includegraphics[width=18cm, height = 10cm]{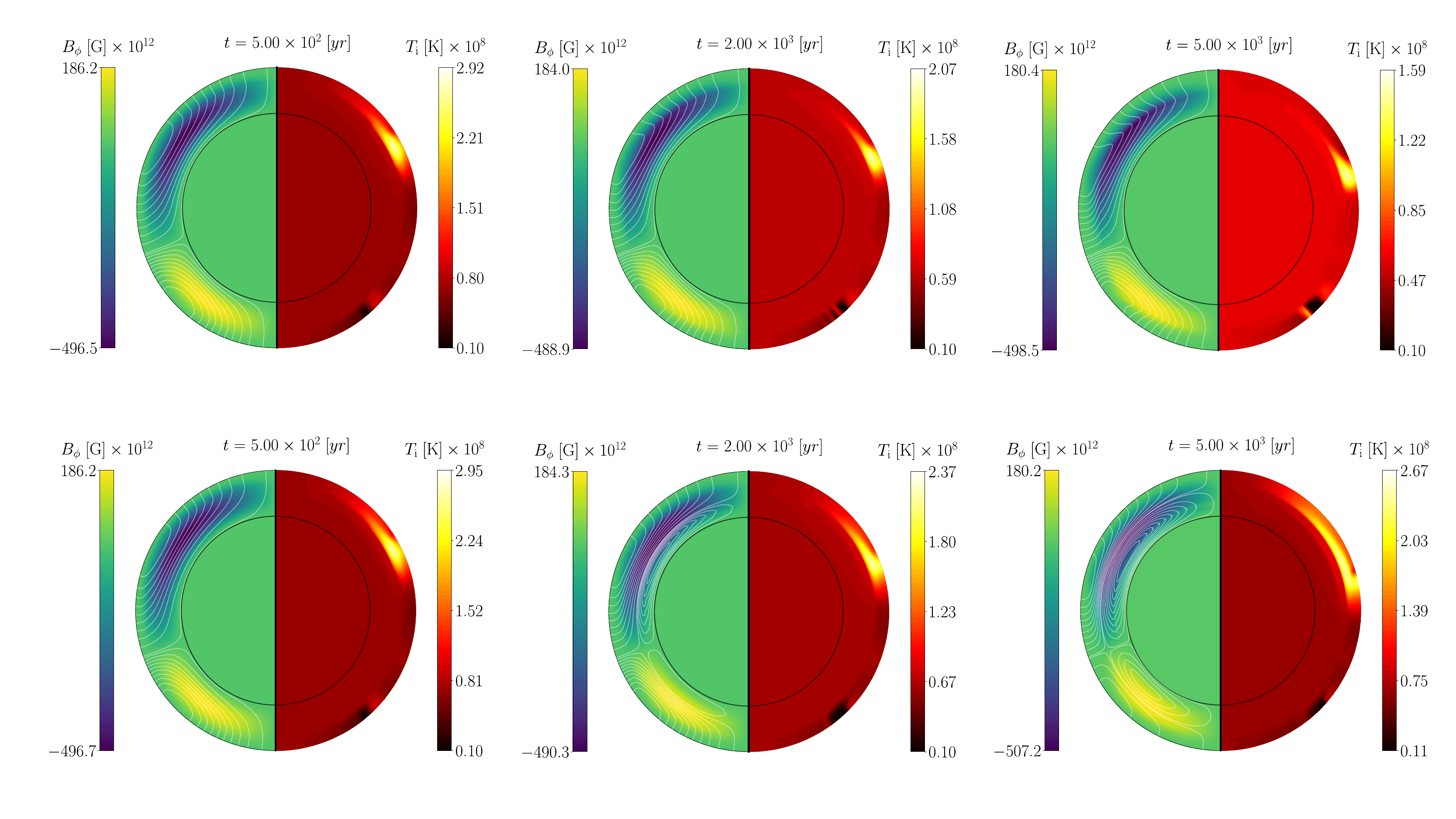}
\caption{As in Figure \ref{fig:pol_tor}, but for the A1m$_{2,2}$ configuration (cf. Table \ref{tab:table_1}).}
\label{fig:multi2_f}
\end{figure*}

We now report the magneto-thermal maps for some of the initial configurations in Table \ref{tab:table_1}.
Figure \ref{fig:pol_tor} studies in detail the magneto-thermal evolution of a star with the initial A1 configuration for $f_a = 5.98 \times 10^{15}$ GeV. The left hemisphere displays the contours of the toroidal field ($\mathbf{B}_\textrm{tor} = B_{\phi}\hat{\phi}$, in units of $10^{12}$ G), and the white lines are the meridional projections of the poloidal magnetic field lines. The right hemisphere reports the redshifted internal temperature $T_{\textrm{i}}$ (in units of $10^8$ K). The comparison between the top and bottom panels (snapshots at $t = 2 \times 10^{3}$ yr) reveals that the dynamo develops a north-south asymmetry for the magnetic field configuration. At $t = 5 \times 10^3$ yr, the bottom panel shows that $B_{\phi}$ reaches maximal values of the order of $10^{15}$ G. In the temperature map, a thick, hot layer forms below the outer envelope (i.e. the outer boundary in the plot) in the southern hemisphere, breaking the equatorial symmetry of $T_{\textrm{i}}$.

We report the magnetic and internal temperature maps for the A1m$_{2,2}$ initial magnetic configuration (cf. Table \ref{tab:table_1}) in Figure \ref{fig:multi2_f}. The initial magnetic energy content of the star is lower than for the A1 configuration, and we use $f_a = 5.98 \times 10^{15}$ GeV. As in Figure \ref{fig:pol_tor}, the top panels in Figure \ref{fig:multi2_f} report the magneto-thermal evolution of models that do not source axions, while the bottom panels display models sourcing axions. The dynamo mechanism produces a rich dynamics. At $t = 5 \times 10^2$ yr, the magnetic configuration is similar for the top and bottom panels. At $t = 2 \times 10^3$ yr, the magnetic field map shows the formation of two closed poloidal loops at the bottom of the crust. At later times ($t = 5 \times 10^3$ yr), the maximal values of the toroidal component of the magnetic field in the bottom and top panels are similar, and the closed loops extend from the bottom to the middle of the crust; the $T_{\textrm{i}}$ map in the bottom panel shows that the crust is hotter with respect to a star not sourcing axions.

\section{\label{sec:Core_fields} Core-extended magnetic fields.}

We now turn to the case of core-extended magnetic fields. The evolution of core-threading magnetic fields is uncertain due to the presence of several fluids in the core regions of NSs, and due to the occurrence of superconducting phases. For proton superconductivity, recent calculations \citep{Wood_2020} predict the coexistence of type-II superconductivity \citep{Baym_1969, Sedrakian_2019}, and type-I superconductivity in mesoscopic regions (smaller than the typical length-scale of the star, but larger than magnetic flux tubes). The former type of superconductivity allows the magnetic field to thread the core, while the latter leads to the expulsion of the magnetic field due to the Meissner effect. Even if the magnetic field is expelled completely from the core region, it is expected to thread the core during the proto-NS stage, when the NS temperature is above the critical temperature for the proton superconductive phase transition.

\begin{figure}
\includegraphics[width=8cm, height = 5.5cm]{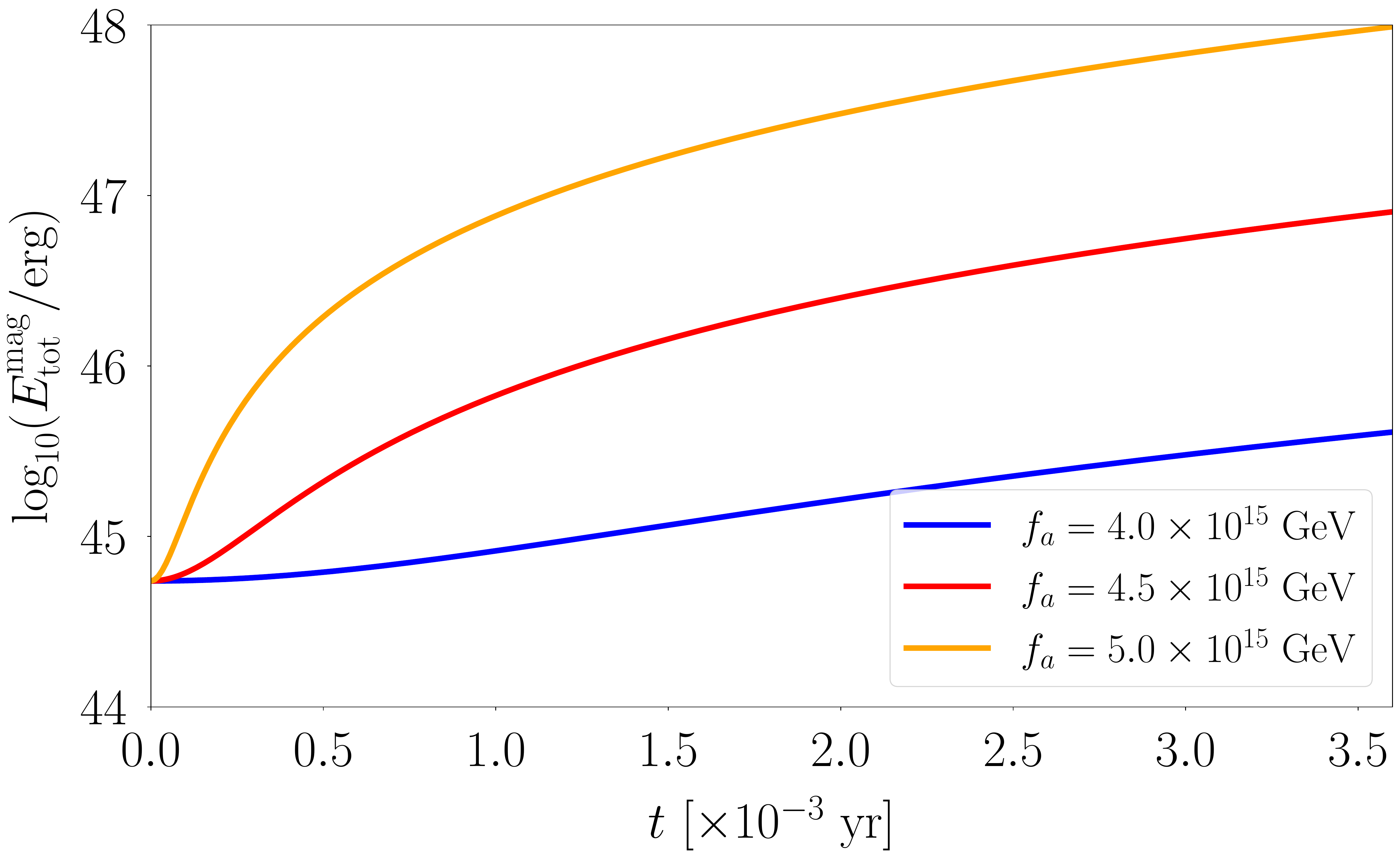}
\caption{\label{fig:Core_extended} Evolution of $E^{\textrm{mag}}_\textrm{tot}$ for the C1 initial magnetic field configuration (core-threading). The dynamo operates on short time-scales ($\approx 10^{-3}$ yr) because the magnetic field is present where $\nabla a$ is far greater than in the crust. }
\end{figure}

In Figure \ref{fig:Core_extended} we consider a core-threading, poloidal-dipolar magnetic field plus a crust-confined, poloidal-dipolar field  (i.e. electric currents are present both in the crust and in the core). Moreover, there is an equatorial torus where the toroidal field is confined (cf. \citep{Vigano_2013, Dehman_2020, Anzuini_2022} for details). For the axion decay constant, we consider the values $4\times 10^{15} \ \leq f_a/\textrm{GeV} \leq 5\times 10^{15}$. In these configurations, the magnetic field is present in the region where the gradient of the axion field $\nabla a$ is considerably larger than in the crust. This affects the time-scale for the growth of the total magnetic energy $E^{\textrm{mag}}_\textrm{tot}$ of the star. For example, in just $t \approx 10^{-3}$ yr, $E^{\textrm{mag}}_\textrm{tot}$ grows to the typical magnetic energy content of magnetars (orange curve). We stop the evolution when $E^{\textrm{mag}}_\textrm{tot}$ attains $E^{\textrm{mag}}_\textrm{tot} \approx 10^{48}$ erg. 

The observable consequences of such fast growth of $E^{\textrm{mag}}_\textrm{tot}$ requires simulations on longer time-scales and in the ultra-strong magnetic field regime. Moreover, since the NS (and in particular the crust) is hot for $t \lesssim 1$ yr, $L_{\gamma}$ is in the range $\gtrsim 10^{36}$ erg/s, and the effect of enhanced Joule heating caused by axions is not visible on the cooling curve on such short time-scales. We leave the study of such configurations on secular time-scales to future work, as it requires simulations in the ultra-strong magnetic field regime, beyond the scope of our work.

\section{\label{sec:Maxwell_equations_setII} \enquote{Standard} equations for axion electrodynamics}
In this section we use the set of Maxwell's equations in the presence of axions widely employed in the literature on direct axion searches \citep{Sikivie_1983, Wilczek_1987, Kim_2019}. Using the same notation as in the main Letter, one has 

\begin{subequations}
   \begin{align}
   &  \nabla \cdot (\mathbf{E} - g_{a\gamma}a\mathbf{B}) = 4\pi\rho_c \\
   \nonumber \\
   & \nabla \times \Big[e^\nu(\mathbf{B}  + g_{a\gamma}a\mathbf{E})\Big] = \frac{\partial (\mathbf{E} - g_{a\gamma}a\mathbf{B})}{c \ \partial t} +  \frac{4\pi e^\nu}{c}\mathbf{J} \ \ \ \ \ \ \ \ \\
   \nonumber \\
   &  \nabla \cdot \mathbf{B} = 0 \\
   \nonumber \\
   & \nabla \times (e^\nu \mathbf{E}) = -\frac{1}{c}\frac{\partial \mathbf{B}}{\partial{t}} \ .
   \end{align}
   \label{eq:Maxwell_k} 
   \end{subequations}
The difference with Eqs. \eqref{eq:VIS_eqs} in the main Letter is that the axion terms appear only in Gauss' law for the electric field and in Amp\`ere's equation. In order to decouple the standard electromagnetic field and the axion contributions, an effective approximation suitable for direct searches has been presented in \citep{Kim_2019}. Assuming that $g_{a\gamma}$ is small, axions perturb the total electromagnetic field, so that

\begin{subequations}
    \begin{align}
    & \mathbf{E} \approx \mathbf{E}_0 + \mathbf{E}_a = \mathbf{E}_0 +  g_{a\gamma}\mathbf{E}_1 \ ,\\
    \nonumber \\
    & \mathbf{B} \approx \mathbf{B}_0 + \mathbf{B}_a = \mathbf{B}_0 + g_{a\gamma}\mathbf{B}_1 \ ,
    \end{align}
    \label{eq:B_field_expansion}
\end{subequations}
where terms proportional to higher powers of $g_{a\gamma}$ are neglected. By substituting the expressions above in Eqs. \eqref{eq:Maxwell_k}, one can solve independently for the background fields $\mathbf{E}_0$ and $\mathbf{B}_0$, and for the perturbations induced by axions, i.e. for $\mathbf{E}_a$ and $\mathbf{B}_a$, the \enquote{reacted} fields \citep{Kim_2019}. The equations for the axion perturbations at first order in $g_{a\gamma}$ read

\begin{subequations}
   \begin{align}
   &\nabla \cdot (\mathbf{E}_a - g_{a\gamma}a\mathbf{B}_0) = 0 \\
   \nonumber \\
   &\nabla \times \big[e^\nu(\mathbf{B}_a + g_{a\gamma}a\mathbf{E}_0)\big] = \frac{\partial (\mathbf{E}_a - g_{a\gamma}a\mathbf{B}_0)}{c \ \partial t} \\
   \nonumber \\
   &\nabla \cdot \mathbf{B}_a = 0 \\
   \nonumber \\
   &\nabla \times (e^\nu \mathbf{E}_a) = -\frac{1}{c}\frac{\partial \mathbf{B}_a}{\partial{t}} \ .
   \end{align}
   \label{eq:Maxwell_k_axion_reacted} 
\end{subequations}
As noted in \citep{Kim_2019}, Eq. (\ref{eq:Maxwell_k_axion_reacted}d) is a generalized version of Faraday's equation which is satisfied by dynamical fields; it allows to determine the evolution of the field $\mathbf{B}_a$, given the induced field $\mathbf{E}_a$.

To determine $\mathbf{E}_a$, we take the curl of Eq. (\ref{eq:Maxwell_k_axion_reacted}d). Neglecting the redshift factors for simplicity, one gets

\begin{eqnarray}
   \nabla^2\mathbf{E}_a = &&     
    \ \frac{1}{c^2}\frac{\partial^2(\mathbf{E}_a - g_{a\gamma}a\mathbf{B}_0)}{\partial t^2} + g_{a\gamma}\nabla[\nabla a \cdot \mathbf{B}_0] -\frac{g_{a\gamma}}{c}\nabla\times\frac{\partial(a\mathbf{E}_0)}{\partial t}  \ .
   \label{eq:axion_waves} 
\end{eqnarray}
As shown in \citep{Kim_2019} (for the case of a haloscope), the solution to $\mathbf{E}_a$ is proportional to $\mathbf{B}_0$; by plugging the $\mathbf{E}_a$ field, solution of Eq. \eqref{eq:axion_waves}, in Maxwell-Faraday equation, the latter acquires a dynamo term, similarly to what found for the induction equation in Eq. \eqref{eq:Maxwell_eq_induction}. 

\begin{figure}
\includegraphics[width=8cm, height = 5.5cm]{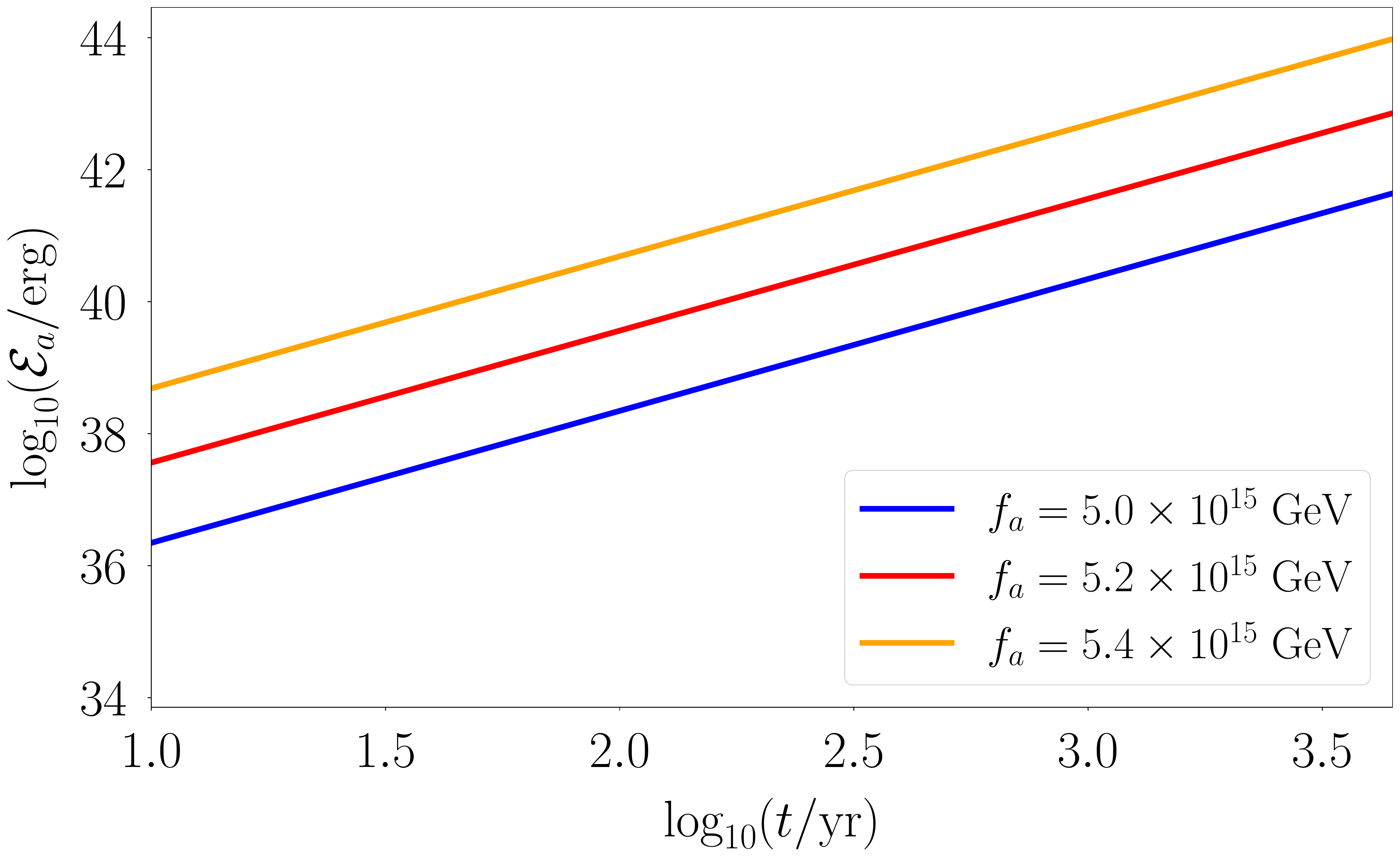}
\caption{\label{fig:ST_dynamo} Total magnetic energy $\mathcal{E}_a$ calculated from the reacted field $\mathbf{B}_a$ versus time, for $f_a = 5.0 \times 10^{15}, 5.2 \times 10^{15}, 5.4 \times 10^{15}$ GeV.}
\end{figure}

We now study the conditions for the activation of the dynamo with the standard Maxwell-axion equations, and simulate the evolution of the total magnetic energy associated with the axion-induced magnetic field $\mathbf{B}_a$, which we denote with $\mathcal{E}_a$. In the following, $\mathcal{E}_0$ stands for the total magnetic energy calculated from the background magnetic field $\mathbf{B}_0$. In general, since the approach adopted in Eqs. \eqref{eq:B_field_expansion} is perturbative, one has to ensure that $\mathbf{B}_a \ll \mathbf{B}_0$ throughout the evolution (and hence $\mathcal{E}_0 \gg \mathcal{E}_a$).

Figure \ref{fig:ST_dynamo} shows the growth of $\mathcal{E}_a$ for the A1 initial magnetic configuration. We use values of $f_a$ such that the axion-induced field $\mathbf{B}_a$ is small compared to the background field $\mathbf{B}_0$ for $t \lesssim 5 \times 10^3$ yr, i.e. $f_a = 5.0 \times 10^{15}, 5.2 \times 10^{15}, 5.4 \times 10^{15}$ GeV. For higher values of $f_a$ (similar to the ones studied in the main Letter), the axion-induced field grows on time-scales shorter than the one considered in Figure \ref{fig:ST_dynamo}, the backreaction on the background field is not negligible anymore for $t \lesssim 5 \times 10^3$ yr, and the perturbative approach in Eqs. \eqref{eq:B_field_expansion} is no longer valid.



\end{document}